\documentstyle[12pt]{article}

\def \be{\begin{equation}}
\def \o{\over}
\def \ee{\end{equation}}
\def \eea{\end{eqnarray}}
\def \bea{\begin{eqnarray}}
\def \l{\label}
\begin{document}
\begin{titlepage}
\vspace*{5mm}
\begin{center}{\Large \bf Electrostatic self--energy in $QED_2$ on curved background}
\end{center}
\begin{center}
\vskip 1cm {\bf H. Mohseni Sadjadi \footnote
{e-mail:amohseni@khayam.ut.ac.ir}}
\vskip 1cm {\it Physics Department, College of Basic Sciences \\
 Shahed University, PO Box 15875-5794} \\
{\it Tehran, Iran }
\end{center}
\vskip 2cm

\begin{abstract}
By considering the vacuum polarization, we study the effects of
geometry on electrostatic self--energy of a test charge near the
black hole horizon and also in regions with strong and weak
curvature in static two dimensional curved backgrounds. We discuss
the relation of ultraviolet behavior of the gauge field propagator
and charge confinement.
\end{abstract}
\begin{center}
\end{center}
\end{titlepage}
\newpage
\section{Introduction}
One of the interesting areas in physics, is the study of behavior
of classical and quantum fields in curved background and
investigating how their properties are affected by the curvature
of the space--time. It also provides a lot of insights into
important problems such as black hole entropy, Hawking radiation,
quantum theory of gravity and so on.

It is well known that the change in the geometry of the space
associated with the gravitational field, deforms electromagnetic
field, inducing a self--force on a point charge at rest in a static
curved space--time\cite{black}.

Also the presence of boundary condition like the boundary
condition at conical singularity produced for example by a cosmic
string or a point mass, alters the electromagnetic field of a
point charge which after subtracting the infinite part, leads to a
finite self--force \cite{cone}.

A renewed investigation has been appeared in this subject in order
to study the upper bound on the entropy of charged object by
requiring the validity of thermodynamics of black--holes. This
problem is studied in \cite{thermo}, for classical black-hole
backgrounds, in the absence of dynamical fermions, i.e.
disregarding vacuum polarization.

Another subject of studies in gauge field theory is the screening
and confinement of charges. The static potential between external
charges, which can be obtained from the Wilson loop expectation
value, carries important information of infrared behavior of gauge
fields which is suggested to be responsible for confinement,
binding the quarks and anti--quarks into $q\bar q$ pairs (infrared
slavery). Because of computational hurdles in four dimensions one
can consider these problems in lower dimensional models, as a
laboratory to study physical effects which can be carried out
to the real world.

In this letter we study the influence of curvature on the
self--energy of static charges, by considering the effect of
vacuum polarization in two dimensional static space--times. We
show that in order to explain the confining behavior of $QED_2$
\cite{shwin} on a curved background, considering ultraviolet
behavior of gauge fields and self--energy of external charges is
necessary, in other words our method of studying confinement
involves the behavior of two point function in the ultraviolet
regime instead of the infrared. To do so We use the criterion
expressing that in the confinement phase the energy of an isolated
quark is infinite. In usual calculations in four dimensional flat
space--time the self--energy of a test charge is infinite and is
subtracted from the potential energy, but on curved space--times
the finite part of self--energy is not a constant and must be
considered in computing the forces\cite{cone}. In two dimensions
this self--energy is related to the Green function of a Sturm--
Liouville type operator at coincident limit and therefore is
analytic. We show that, for $QED_2$, the self--force can prohibit
a single charge to be in some region of the space unless it is
coupled to another opposite test charge, forming mesonic $q{\bar
q}$ structure. We also obtain electrostatic self--force using the
heat kernel method up to the adiabatic order four.
\section{ Geometrical effects on charge confinement and mesonic structure}
A general static two dimensional surface can be described by the
metric \be \l{1} ds^2={\sqrt g}(x)(dt^2-dx^2),  \ee where ${\sqrt
g}(x)$ is the conformal factor.\footnote{ Euclidean version of
this space--time is $ds^2={\sqrt g}(x)(dt^2+dx^2)$, whose one of
the geodesics is the straight line parallel to the $x$ axis.} On
this space--time, $QED_2$ consisting of charged matter field
interacting with an abelian gauge field in two dimensions is
described by Lagrangian \be \l{2} L={\sqrt g}(x)
(\psi^{\dagger}\gamma^{\mu}(\nabla_\mu -ieA_\mu)\psi ) +{1\over
2{\sqrt g}(x)}F^2), \ee where $\gamma^\mu $ are the curved space
counterparts of Dirac gamma matrices. $\nabla_{\mu}$ is the
covariant derivative, including the spin connection, acting on
fermionic fields. $e$ is the charge of dynamical fermions. The
dual field strength $F$, is described through $F=\epsilon ^{\mu
\nu}
\partial_{\mu}A_{\nu}$,
where $ \epsilon^{\mu \nu}=\epsilon_{\mu \nu}$ and $
\epsilon^{01}=-\epsilon_{10}=1$.
 By integrating out matter fields one can obtain one loop
 effective action for the gauge field \cite{schwcurved}
 \be \l{3}
 L_{eff.}={1\o 2{\sqrt g}(x)}F^2 +{\mu^2\o 2}F{1\o {\partial^2}}F,
\ee where $\mu ={e\o {\sqrt \pi}}$. In static case and using the
Coulomb gauge $A_1=0$ this Lagrangian reduces to \be \l{4}
 L_{eff.}=
{1\o 2{\sqrt g}(x)}({dA_0\over dx})^2 +{\mu^2\o 2}A_0^2. \ee Hence
in the presence of vacuum polarization,  the gauge field has
gained a mass via a peculiar two dimensional version of Higgs phenomenon.
As a consequence, one may expect the replacement of
the Coulomb force by a finite range force. We introduce two static
opposite charges located at $x=a$ and $x=b$, described by the
covariantly conserved current \be \l{5} J^0(x) ={e'\o{\sqrt
g}(x)}(\delta (x-b) -\delta (x-a)), \ \ \ \ J^1=0. \ee  The gauge
field's equation of motion is \be \l{6} {d\o dx}{1\o {\sqrt
g}(x)}{dA_0\o dx}- \mu^2 A_0= e'(\delta (x-b)-\delta (x-a)). \ee
The Green function of the elliptic operator ${d\o dx} {1\o {\sqrt
g}(x)}{d\o dx}-\mu^2$ satisfies \be \l{7}  ({d\o dx}{1\o {\sqrt
g}(x)}{d\o dx}- \mu^2)G(x,x')=\delta (x,x'). \ee  In terms of
$G(x,x')$ the energy of external charges is obtained \be \l{8}
E=\int T^0_0dx=-\int L_{eff.}dx=-{{e'^2}\o
2}[G(a,a)+G(b,b)-2G(a,b)]. \ee This is the energy measured by an
observer whose velocity $u^\mu =(g^{-1/4}(x),0)$ is parallel to
the direction of the time--like killing vector of the space--time
and ${-e'^2\o 2}G(x,x)$ is the self--energy of a static point
charge located at $x$.

On flat surfaces this change of energy
is
\be \l{p}
E={e'^2\over 2\mu}(1-e^{-\mu|b-a|}).
\ee
This can be obtained by using the gauge fields Green's function
\be \l{o}
G(x,x')=-{1\over 2\mu}e^{-\mu|x-x'|}
\ee
Note that $-{e'^2\over 2}G(x,x)$, in contrast to higher dimensions
has a finite value. This is due to finiteness of Helmholtz Green's function
in coincident limit in one dimension.
For distant charges as a result of screening,  interaction
term becomes zero, and  $E$
tends to self--energy of test charges, which in
flat case is $E_{|b-a|\to \infty}={e'^2\over 2\mu}$ \cite{jac}. Note that
although this self--energy is finite, but it is a constant therefore
the self--force is zero.  As we will show on a curved surface, the Green function
of the Sturm--Liouville operator (\ref {7}) is an analytic
function at coincident limit and self--force
becomes position dependent.

Besides, the role of dynamical charge is affected by the presence of the curvature
\be \l{9} R(x)={1\over
{\sqrt g}(x) }{d\over dx}{1\over {\sqrt g}(x)}{d\over dx}{\sqrt
g}(x), \ee
To see this we write the equation (\ref{7}) as \be
\l{10} {1\over \sqrt g(x)}{d\over dx}{1\over \sqrt g(x)}{d\over
dx}\sqrt g(x) {\tilde G(x,x')}-\mu^2 {\tilde G(x,x')}={\delta
(x,x')\over {\sqrt g(x)}}, \ee where $G(x,x')=\sqrt g(x){\tilde
G(x,x')}$. This equation can be rewritten as \be
\l{11}(R(x)-\mu^2){\tilde G(x,x')}+{1\over g(x)}({d{\sqrt
g}(x)\over dx})({d{\tilde G(x,x')}\over dx})+{1\over {\sqrt
g}(x)}{d^2{\tilde G(x,x')}\over dx^2}={\delta(x,x')\over
{\sqrt{g}}(x)}, \ee which is equivalent to \be
\l{12}(R(x)-\mu^2)G(x,x')+{1\over {\sqrt g}(x)}({d{\sqrt
g}(x)\over dx})({d\over dx}{1\over {\sqrt g}(x)}
G(x,x'))+{d^2\over dx^2} ({G(x,x')\over {\sqrt
g}(x)})=\delta(x,x'). \ee Hence in the strong curvature limit
$|R(x)|\gg \mu^2$, the Green function is approximately unaffected
by dynamical fermions. In other words if we assume the same
boundary condition for the gauge field in the presence and absence
of dynamical fermions, the vacuum polarization, in contrast to the flat case,
doesn't change the energy and
confining phase of the system.

To elucidate this
subject and to emphasize how the ultraviolet behavior of $QED_2$
on curved space--time is concerned in confinement of test charges
let us give an example. Consider the following space--time in
conformal coordinates \be \l{13} ds^2={{dt^2-dx^2}\over x^m}
;\;\;\;\;x> 0 \ee where $m=2-1/(k+1)$; $k\neq 0,-{1/2}$, this is
one of the classical solutions of two dimensional scale invariant
gravity \cite{scaleinv} with a curvature singularity. Here we
assume that this is only a classical background for $QED_2$. The
homogenous solutions $G_h(x)$ of the equation (\ref{7}) satisfy
\be \l{14}{d\over dx}x^m{d\over dx}G_h(x)-\mu^2 G_h(x)=0.\ee

For $m\neq 2$, by defining $z\equiv x^{1-{m\over 2}}$ and
$G_h(x)\equiv x^{{1-m}\over 2}u$, we obtain \be
\l{15}z^2{d^2u\over dz^2}+z{du\over dz}-({(m-1)^2\over
(m-2)^2}+{4\mu^2 z^2\over(m-2)^2})u=0. \ee Hence
\begin{eqnarray}\l{16}
G _h(x)=  x^{{1-m}\over 2}\left\{ \begin{array}{ll}
 I_{|{{m-1}\over {m-2}}|}(\sqrt{4\mu^2\over
(2-m)^2}x^{1-{m\over 2}})& \\
K_{|{{m-1}\over {m-2}}|}(\sqrt{4\mu^2\over
(2-m)^2}x^{1-{m\over 2}}). & \\
\end{array} \right.
\end{eqnarray}
Therefore the Green function is
\begin{eqnarray}\l{17}
 G(x,x')=-{2 \over
|m-2|}(x_{<}x_{>})^{1-m\over 2}\left\{
\begin{array}{ll}
I_{{{m-1}\over {m-2}}}({2\mu\over {m-2}}x_{>}^{1-{m\over 2}})
K_{{{m-1}\over {m-2}}}({2\mu\over {m-2}}x_{<}^{1-{m\over 2}})&
\textrm{if $m>2$}\\
I_{|{{m-1}\over {m-2}}|}({2\mu\over {2-m}}x_{<}^{1-{m\over 2}})
K_{|{{m-1}\over {m-2}}|}({2\mu\over {2-m}}x_{>}^{1-{m\over 2}})&
\textrm{if $m<2$},
\end{array} \right.
\end{eqnarray} where $x_{>(<)}$ is the bigger (smaller) of $x,x'$ and $I$, $K$
are modified Bessel functions.
This Green function Satisfies Dirichlet boundary condition at
$x=0$ and at $x=\infty$. At the coincidence limit the Green
function is
 \be \l{18}
G(x,x)=-{2 \over |m-2|}x^{1-m}I_{|{{m-1}\over
{m-2}}|}\big({2\mu\over {|m-2|}}x^{1-{m\over 2}}\big )
K_{|{{m-1}\over {m-2}}|}({2\mu\over |m-2|}x^{1-{m\over 2}}).
 \ee
 In terms of the scalar curvature $R(x)={m\over x^{2-m}}$, this
 relation becomes
 \be \l{19}
 G(x,x)=-{2 \over |m-2|}\Big({m\over R(x)}\Big)^{{1-m}\over {2-m}}I_{|{{m-1}\over
{m-2}}|}\Big({2\mu\over {|m-2|}}({R(x)\over m})^{-{1\over 2}}\Big)
K_{|{{m-1}\over {m-2}}|}\Big({2\mu\over {|m-2|}}({R(x)\over
m})^{-{1\over 2}}\Big).
 \ee
 In the strong curvature limit, or in regions where ($|R(x)|\gg \mu^2$), by considering
 the asymptotic behavior of Bessel functions
 this relation becomes
 \be \l{20}
 \lim_{R\to \infty}G(x,x)=-|{k+1\over k}|({R(x)\over m})^k,
 \ee
 which is $\mu$ independent as anticipated. $k$ is defined after the equation (\ref{13}).
 In regions where the curvature is weak $|R(x)|\ll \mu^2$, we
 have
 \be \l{21}
 G(x,x)=-{1\over 2\mu}({R(x)\over m})^{m\over 2(2-m)}.
\ee
Note that for $m=0$, $ G(x,x)=-{1\over 2\mu}$, which is the same as the result on
a flat uncurved surface \cite{jac}.

 Now we show the relation of electrostatic self--energy and
confinement.
 $G(x,x)$ in (\ref{19}) is an analytic non--constant function for $x>0$
 (which may become very large in some regions, signaling, as we will show, a confining
 situation), and
 therefore we expect that the charge feels an electrostatic self--force, affected
 by the geometry of the space--time as well as the boundary condition imposed
 on the gauge fields.

We use the fact that in the confinement phase the energy of
an isolated quark is infinite.

Assume that $m>2$; the energy needed to locate a
single charge $e'$ in the region $(R(x)\neq 0)\ll \mu^2$ or $x\simeq 0$
(in the coordinate (\ref{13})), is \be \l{22}
E_{self.}(x)={{e'}^2\over 4\mu}({R(x)\over m})^{m\over 2(2-m)},\ee
which tends to infinity. In other words there is a great repulsive
force on an external charge near $x=0$ prohibiting to have single
charges in this region, or the energy of an isolated
charge in this region is very large. 
The same procedure occurs in the region
$x\simeq 0$, for $m<2$ (when $|R(x)|\to \infty$). So in these
regions following the equation (\ref{8}) only charges forming
mesonic structure may survive. These two opposite charges must be
near together in order to obtain a finite energy for the system. 
Our criterion for confinement is not based on the
behavior of the energy in the infrared (where the geodesic
distance of external charges tends to infinity).

In $QED_4$ the dielectric constant of vacuum is larger than unity
as a result of the screening effect due to vacuum polarization. If
instead one consider a case in which the dielectric constant of
the vacuum vanishes then due to antiscreening effect the energy of
the system becomes infinite unless we add another opposite test
charge to the system then this fictitious system is confining. A
similar phenomenon  occurs in dual bag model. In that case quarks
and anti--quarks must form mesonic structure in order to avoid
divergences in static potential, i.e only color singlets have
finite energy because the divergence term appears as a multiplier
of total external charges \cite{dual}. In our model the role of
dielectric constant in confining phase is played by the metric
components (see equation (\ref{2})). To find some relations
between geometry and permittivity see \cite{diel}.

The repulsive forces on single charges besides the curvature of
the space--time is related to the boundary conditions imposed on
the gauge fields in defining the vacuum of the system, for example in equation
(\ref{24}),
although $R$ is small but we have a great repulsive force. These
repulsive forces may also be arisen on a surface with constant
curvature (adS or dS space--times, obtained for example by taking
$m=2$ in the previous example)\cite{sad}. Also
Maxwell field theory (disregarding vacuum polarization) in 2+1
dimensional conical space--times, despite the null curvature of
the manifold exhibits a repulsive force on charges \cite{cone}.

In the previous example we considered a space--time with a naked singularity,
in this part we study the electrostatic self-- energy on a black hole
background.

For Maxwell theory on a four dimensional Schwarzschild black-hole
a test charge near the horizon is repelled by an image charge
inside the horizon. In these cases one must subtract the infinite
parts to obtain a renormalized Green function or the finite part
of self--energy. On a two dimensional static space--time, in
contrast to the Maxwell theory in three and four dimensions, as we
noticed (after equation (\ref{o})) and
will discuss later,  electrostatic self--energy in $QED_2$ is a well
defined function.

In  Schwarzschild coordinate, we consider a non--extremal two
dimensional static black hole described by the metric \be
\l{23}ds^2=f(r)dt^2-{1\over f(r)}dr^2. \ee At the horizon $r=h$,
$f(h)=0$. In this coordinate the equation (\ref{7}) becomes \be
\l{24}\Big(f(r)({d^2\over
dr^2})-\mu^2\Big)G(r,r')=f(r)\delta(r,r').\ee Near the (bifurcate)
horizon, i.e. $r \simeq h$, $r>h$, we have $f(r)=\kappa (r-h)$,
where $2\kappa$ denotes the surface--gravity. Assuming the gauge
field tends to zero at infinity and is well behaved at the
horizon, the two point function of the gauge field becomes \be
\l{25} G(r,r')=2(r_{>}-h)^{1\over 2}(r_{<}-h)^{1\over
2}K_1(2\mu\sqrt{{r_>-h\over \kappa}})I_1(2\mu\sqrt{{r_<-h\over
\kappa}}),\ee where $r_{<(>)}$ is the smaller (bigger) of $r$ and
$r'$.

To obtain the effect of the gravitational field on the
electrostatic self--interaction, we use the global method used in
\cite{will}. If in a free falling coordinates the work $\delta W$ is
needed to displace the charge slowly by a distance $\delta r$,
then this energy computed at asymptotic infinity, by considering
the gravitational red--shift, will be (the space--time is flat at
infinity) \be \l{red} \delta E={\sqrt f(r)}\delta W. \ee Using the
total mass variation law of Carter \cite{cart}, and assuming that
the metric is unperturbed by the presence of the charges
\cite{will}, we arrive again to the equation (\ref{8}). Note that
the integral must be taken over $T_0^0$ from the horizon to
infinity. The self--energy of a test charge near the horizon is
then \be \l{26} E_{self.}(x)=e'^2(r-h)K_1\Big(2\mu \sqrt
{{r-h}\over \kappa}\Big)I_1\Big(2\mu \sqrt {{r-h}\over
\kappa}\Big).\ee In contrast to the four dimensional case, there
is an attractive force on the test charge near the horizon, which
by considering the asymptotic behavior of Bessel functions is
independent of vacuum polarization: this attractive force may be
related to an image charge inside the horizon, when these charges
are near together, that is near the horizon, the effect of vacuum
polarization may be disregarded (self--energy is independent of
$\mu$). At the horizon the self--energy is zero, and in contrast
to the previous example ultraviolet behavior of the Green function
doesn't lead to charge confinement. Vanishing of self--energy may
be understood as follows: Instead of the black--hole horizon one
can consider an image charge inside the black--hole. Then the
self--energy of the test charge is \be \l{test} E={-{e}'^2\over
2}({\tilde G}(r,r)-{\tilde G}(r,r')), \ee where $r'$ is the
location of the image charge $-e'$,  and ${\tilde G}$ is the
Green's function which does not satisfy the Dirichlet boundary
condition. In the limit $r=r'=h$ we obtain $E=0$.

Besides,
if we assume $r\rightarrow \infty$, the interaction energy between
the charge and its image becomes zero, and we obtain $E={e'^2\over
4\mu}$, which is the self--energy of a test charge on a flat
surface. This can be seen explicitly by setting $f(r\rightarrow
\infty)=1$.

\section{Heat kernel expansion of electrostatic self--energy}

In this part we study the short distance behavior of Green
function $G(x,x')$, using the heat kernel of positive elliptic
operator $O:=-{d\o dx} {1\o \sqrt g}{d\o dx}+\mu^2$.
This method can be used for slowly varying metrics. We write the
heat kernel in the form \be \l{27} h(\tau
;x,x')=\sum_{n=0}^{\infty}{\tau^{(n-{1\over 2})}\over
\sqrt{4\pi}}\exp({-\sigma \over 2\tau}-\mu^2 \tau)a_n(x,x'),\ee
which satisfies \be \l{28} Oh(\tau;x,x')+{\partial
h(\tau;x,x')\over
\partial \tau}=0,\ee where $\sigma={1\over 2}\mid
\int_{x'}^{x}g^{1\over 4}(y)dy\mid^2$, is one half of the square
of geodesic distance between $(t,x)$ and $(t,x')$ and $\tau$ is
the proper--time parameter. $G(x,x')$ is given by \be
\l{29}G(x,x')=-\int_{0}^{\infty}h(\tau;x,x')d\tau,\ee provided \be
\l{30} h(0;x,x')=\delta( x,x').\ee Therefore \be
\l{31}G(x,x')=-{1\over \sqrt \pi} \sum_{n=0}^\infty ({\sigma \over
2\mu^2})^{({n\over 2}+{1\over 4})}K_{n+{1\over 2}}(\mu
\sqrt{2\sigma})a_{n}(x,x') \ee and for $\sigma=0$,\be \l{32}
G(x,x)=-{1\over 2\sqrt \pi} \sum_{n=0}^{\infty}(\mu^2)^{-n-{1\over
2}}\Gamma(n+{1\over 2})a_n(x,x),\ee which is regular.

Under a scale transformation parameterized by the positive number
$\lambda$,
\begin{eqnarray}\l{33}
&&{\sqrt g}(x)\rightarrow \lambda {\sqrt g}(x)
\nonumber \\
&& R(x)\rightarrow {1\over \lambda}R(x),
\end{eqnarray}\
the Green function becomes
 $G_{\lambda}(\mu^2)=\lambda G(\mu^2
\lambda)$ ${}$or${}$ $ {1\over \lambda} G_{\lambda}({ \mu^2\over
\lambda})= G(\mu^2)$. We have written the $\mu$ dependence of $G$
explicitly. Using (\ref{32} ), we obtain \be \l{34}
a_{n}(\lambda)={a_n(\lambda=1)\over{\lambda^{n-{1\over 2}}}}. \ee
Hence As a polynomial, $a_n$ consists only of $mth$ power of $g$
(including also its derivatives), where $m={1\over 4}-{n\over 2}$.
In $a_n$, the order of derivatives is $2n$. For example as we will
see, in $a_1(x,x)$, only the terms $g^{-{9\over 4}}(x)g'^2(x)$ and
$g^{-{5\over 4}}(x)g''(x)$ are present.

In order to obtain heat--kernel coefficients we use the relations
\be \l{35}[\sigma]=[\sigma']=0,{}[\sigma^{(2)}]={\sqrt g}(x),{}
[\sigma^{(3)}]={3g'\over 4{\sqrt g}}(x). \ee We have shown $\sigma
(x,x)$ by $[\sigma]$ and $\prime$ denotes the first derivative and
$(n)$ the nth
 derivative with respect to $x$.

By solving the equation (\ref{28}) for the Seeley coefficients, we
obtain a recursion relation
\begin{eqnarray} \l{36}&-&{1\over 2}g^{-{3\over 2}}(x)g'(x)a'_{n}(x,x')+g^{-{1\over
2}}(x)a^{(2)}_n(x,x')-(n+1)a_{n+1}(x,x'
)+ {} \nonumber \\
& & {} {1\over 8}g'(x)\sigma'(x,x')g^{-{3\over
2}}(x)a_{n+1}(x,x')-g^{-{1\over 2}}\sigma'(x,x')a'_{n+1}(x,x')=0.
\end{eqnarray}
For $n<0$, $a_{n<0}=0$. In order to satisfy (\ref{30}) we must
have $a_{0}(x,x')=g^{1\over 8}(x)g^{1\over 8}(x')$. Taking the
diagonal value of (\ref{36}) yields \be \l{37}-{1\over
2}g^{-{3\over 2}}(x)g'(x)[a'_{n}]+g^{-{1\over
2}}(x)[a^{(2)}_n]-(n+1)[a_{n+1}]=0.\ee

For $n=0$ \be \l{38}[a_1]=-{1\over 2}g^{-{3\over
2}}(x)g'(x)[a'_0]+g^{-{1\over2}}[a_0^{(2)}], \ee hence \be \l{39}
[a_1]=-{11\over 64}g^{-{9\over 4}}(x){g'}^2(x)+{1\over
8}g^{-{5\over 4}}(x)g^{(2)}(x).\ee For $[a_2]$ we require the
diagonal part of $a_1$  derivatives: $[a_1']$ and $[a_1^{(2)}]$.
Differentiating (\ref{36}) with respect to $x$ gives
\begin{eqnarray} \l{40}& &\Big(-{1\over 2}g^{-{3\over
2}}(x)g^{(2)}(x)+{3\over 4}g^{-{5\over
2}}{g'}^2(x)\Big)[a'_n]-g^{-{3\over
2}}(x)g'(x)[a_n^{(2)}]+g^{-{1\over 2}}[a_n^{(3)}]
-\nonumber \\
& & (n+2)[a'_{n+1}]+{1\over 8}g^{-1}(x)g'(x)[a_{n+1}]=0.
\end{eqnarray}
For $n=0$ \be \l{41} [a'_1]=-{1\over 4}g^{-{9\over
4}}(x)g'(x)g^{(2)}(x)+{99\over 512}g^{-{13\over 4}}{g'}^3+{1\over
16}g^{-{5\over 4}}(x)g^{(3)}(x).\ee Another differentiation of
(\ref{36}) with respect to $x$ in the limit $x\rightarrow x'$
leads to the following equation
\begin{eqnarray}\l{42}
&& \Big(-{1\over 2}g^{-{3\over 2}}(x)g^{(3)}(x)+{9\over
4}g^{-{5\over 2}}(x)g'(x)g^{(2)}(x)-{15\over 8}g^{-{7\over
2}}(x){g'}^3(x)\Big)[a'_n]+\nonumber \\
&&\Big(-{3\over 2}g^{(2)}(x)g^{-{3\over 2}}(x)+{9\over
4}{g'}^2(x)g^{-{5\over 2}}(x)\Big)[a_n^{(2)}]+g^{-{1\over 2}}[a^{(4)}_n]
-{3\over 2}g'(x)g^{-{3\over 2}}(x)[a_n^{(3)}]+\nonumber \\
&&\Big({1\over 4}g^{(2)}(x)g^{-1}(x)-{9\over
32}{g'}^2(x)g^{-2}(x)\Big)[a_{n+1}]-{1\over
2}g'(x)g^{-1}(x)[a'_{n+1}]-\nonumber \\
&&(n+3)[a^{(2)}_{n+1}]=0.
\end{eqnarray}
Therefore we find \begin{eqnarray}\l{43}&& [a_1^{(2)}]={-{23\over
96}}g^{-{9\over 4}}(x)g'(x)g^{(3)}(x)+{1\over 24}g^{-{5\over
4}}(x)g^{(4)}(x)-{31\over 192}g^{-{9\over
4}}(x){g^{(2)}}^2(x)-\nonumber \\
& & {1947\over 4096}g^{-{17\over 4}}(x){g'}^4(x)+{213\over
256}g^{-{13\over 4}}(x){g'}^2(x)g^{(2)}(x).\end{eqnarray}
Equations (\ref{37}), (\ref{41} ), (\ref{43}) yield
\begin{eqnarray}\l{44}
[a_2]=&&{245\over 512}g^{-{15\over
4}}(x){g'}^2(x)g^{(2)}(x)-{2343\over 8192}g^{-{19\over
4}}(x){g'}^4(x)-{26\over 192}g^{-{11\over
4}}(x)g'(x)g^{(3)}(x)-\nonumber \\
&&{31\over 384}g^{-{11\over 4}}(x){g^{(2)}}^2(x) +{1\over
48}g^{-{7\over 4}}(x)g^{(4)}(x).
\end{eqnarray}
One can continue this method to obtain other $[a_n]$.
Heat kernel
coefficient can be expressed in terms of the scalar curvature
$R={1\over 2}g^{-{3\over 2}}(x)g^{(2)}(x)-{1\over 2}g^{-{5\over
2}}(x)g'^2(x)$ and $\kappa(x)={1\over 2}g'(x)g^{-{1}}(x)$ (at the
horizon of a black--hole, $2\kappa$ is the surface --gravity).
\begin{eqnarray}\l{45} &&[a_1]={1\over 4}g^{1\over 4}(x)R(x)-{3\over
16}g^{-{1\over
4}}(x)\kappa^2(x)\nonumber \\
&& [a_2]={1\over 24}g^{-{1\over 4}}(x)R^{(2)}(x)-{1\over
8}g^{-{1\over
4}}(x)\kappa (x)R'(x)-{1\over 32}g^{1\over 4}(x)R^2(x)+\nonumber \\
&&{23\over 192}g^{-{1\over 4}}(x)\kappa^2(x)R(x)-{23\over
512}g^{-{3\over 4}}(x)\kappa^4(x).
\end{eqnarray}

Hence the self--energy of the charge $e'$ up to the fourth
adiabatic order is
\begin{eqnarray}\l{46}
E_{self.}(x)&=&{e'^2\over 4}g^{-{1\over 4}}(x) \Big
[\mu^{-1}g^{1\over 2}(x)+{1\over 8}\mu^{-3}\Big(g^{1\over
2}(x)R(x)-{3\over 4}\kappa^2(x)\Big)+\nonumber \\
&& {3\over 32}\mu^{-5}\Big({1\over 3}R^{(2)}(x)-
\kappa (x)R'(x)-{1\over 4}g^{1\over 2}(x)R^2(x)+\nonumber \\
&&{23\over 24}\kappa^2(x)R(x)-{23\over
64}g^{-1}(x)\kappa^4(x)\Big)\Big].
\end{eqnarray}

Note heat kernel expansion like the WKB method can not be applied
at the horizon $h$ or the turning point,  i.e. where $g(h)=0$. To
obtain an expression for the self--energy near the horizon one can
expand the metric and follows the steps after the equation
(\ref{23}). However the expansion (\ref{46}), is consistent with
the asymptotic behavior of electrostatic self--energy obtained
near the horizon. For example Considering asymptotic behavior of
Bessel function for large arguments in equation (\ref{26}) gives
the first term in equation (\ref{46}). This first term is also the
same as the result obtained for the small curvature limit
(\ref{21}). For $\sqrt{g}=1$, (\ref{46}) is reduced to the flat
case result $E_{self.}={e'^2\over 4}$ \cite{jac}. 

\vskip 1cm
\begin{thebibliography}{99}
\bibitem{black}B. l\'{e}aut\'{e} and B. Linet, Int. J. Theor. Phys. {\bf 22}, 67 (1983).
\bibitem{cone}T. Souradeep, V. Sahni, Phys. Rev. {\bf D 46}, 1616 (1992).
\bibitem{thermo}S. Hod, Phys. Rev. {\bf D 61}, 24023 (2000);
B. Linet, Gen. Rel. Grav. {\bf 31}, 1609 (1999).
\bibitem{schwcurved}J. Barcelos-Neto and A. Das, Phys. Rev. {\bf D
33}, 2262 (1986).
\bibitem{shwin} J. Schwinger, Phys. Rev. {\bf 128}, 2425 (1962).
\bibitem{jac}D. J. Gross, I. R. Klebanov, A. V. Matystin, A. V. Smilga, Nucl. Phys. {\bf B 461}
,109 (1996).
\bibitem{scaleinv}S. Mignemi, H.-J. Schmidt, Class. Quant. Grav. {\bf 12}, 849 (1995).
\bibitem{dual}M. Baker, James S. Ball and F. Zachariasen, Phys. Rep.
{\bf 209}, 73 (1991).
\bibitem{diel} R. Schutzhold, G. Plunien, G. Soff, Phys. Rev. Lett.
{\bf 88}, 061101 (2002).
\bibitem{sad}M. Alimohammadi, H. Mohseni Sadjadi, Phys. Rev. {\bf D 63}, 105018 (2001)
; H. Mohseni Sadjadi, Kh. Saaidi, Phys. Rev. {\bf D 63}, 65009
(2001); H. Mohseni Sadjadi, M. Alimohammadi, Int. J. Mod. Phys.
{\bf A 16}, 1631 (2001).
\bibitem{cart} B. Carter, {\it{General Relativity: An Einstein Centenary Survey,
S. W. Hawking and W. Israel (New York: Cambridge university Press 1979)}}.
\bibitem{will}A. G. Smith, C. M. Will, Phys. Rev. {\bf D 22}, 1276 (1980).
\end{thebibliography}
\end{document}